\documentclass[journal]{IEEEtran}
\usepackage{cite}

\usepackage{graphicx}
\ifCLASSINFOpdf

\else
\fi
\usepackage{color,xcolor}
\usepackage{array}
\usepackage{multirow}
\usepackage{float}

\definecolor{cblue}{RGB}{0,0,0}

\begin{document}
%
\title{Multi-frame Joint Enhancement for Early Interlaced Videos}
%
%
%

\author{Yang Zhao, Yanbo Ma, Yuan Chen, Wei Jia*, Ronggang Wang, and Xiaoping Liu
\thanks{
This work is supported by grants from the National Natural Science Foundation of China (Nos. 61972129, 62076086),  the Key Research and Development Program in Anhui Province (No. 202004d07020008), and Shenzhen Research Projects of JCYJ20180503182128089 and 201806080921419290. W. Jia is the corresponding author.
	
Y. Zhao, Y. Ma, Y. Chen, W. Jia, and X. Liu are with the Key Laboratory of Knowledge Engineering with Big Data, Ministry of Education, Hefei University of Technology, and also with the School of Computer Science and Information Engineering, Hefei University of Technology, Hefei 230009, China (e-mail:yzhao@hfut.edu.cn; ybma@mail.hfut.edu.cn; ychen@mail.hfut.edu.cn; jiawei@hfut.edu.cn; lxp@hfut.edu.cn).

R. Wang is with the School of Electronic and Computer
Engineering, Peking University Shenzhen Graduate School, 2199 Lishui Road,
Shenzhen 518055, China (e-mail: rgwang@pkusz.edu.cn).

Y. Zhao, R. Wang are also with Peng Cheng Laboratory,
Shenzhen 518000, China.}
}

\maketitle

\begin{abstract}
Early interlaced videos usually contain multiple and interlacing and complex compression artifacts, which significantly reduce the visual quality. Although the high-definition reconstruction technology for early videos has made great progress in recent years, related research on deinterlacing is still lacking. Traditional methods mainly focus on simple interlacing mechanism, and cannot deal with the complex artifacts in real-world early videos. Recent interlaced video reconstruction deep deinterlacing models only focus on single frame, while neglecting important temporal information. Therefore, this paper proposes a multiframe deinterlacing network joint enhancement network for early interlaced videos that consists of three modules, i.e., spatial vertical interpolation module, temporal alignment and fusion module, and final refinement module. The proposed method can effectively remove the complex artifacts in early videos by using temporal redundancy of multi-fields. Experimental results demonstrate that the proposed method can recover high quality results for both synthetic dataset and real-world early interlaced videos.
\end{abstract}


\IEEEpeerreviewmaketitle

\section{Introduction}
\IEEEPARstart{I}{nterlaced}  scanning mechanisms have been widely used in early TV broadcasting systems (\textit{e.g.}, NTSC, PAL, and SECAM). Different from the current progressive scanning method, the odd and even rows of pixels in interlaced frame are scanned from two different frames, namely the odd and the even fields (as shown in Fig. 1). In this way, the high frame rate and transmission bandwidth of the video display have been well balanced. Unfortunately, two fields captured at different time instances may cause displacement differences, which cannot be perfectly aligned in space. Hence, as illustrated in Fig. 2, obvious comb artifacts can be observed in the interlaced frames, especially in the case of large motion.

{\color{cblue} With the development of high-definition digital display technology, progressive scanning technology has become mainstream, and thus the researches on deinterlacing has received little attention. However, there are many precious early videos that can arouse the resonance of audiences. At the same time, the risk and cost of remaking classic early videos are lower than that of making new ones. Therefore, in recent years, the reconstruction of early videos has received more and more attentions. Unfortunately, early videos usually contain various artifacts during the processes of digitization, compression, transmission and storage. For example, interlaced artifacts in early interlaced videos are often mixed with various noises. Traditional de-interlacing methods are only designed for deinterlacing task, which  cannot solve such complex artifacts. In summary, the main purpose of this paper is to design an effective joint enhancement model to reconstruct the real world early interlaced videos. 
\begin{figure}[!t]
	\setlength{\abovecaptionskip}{0.2cm}
	\setlength{\belowcaptionskip}{0.2cm}
	\centering
	\includegraphics[height=8cm,width=0.48\textwidth]{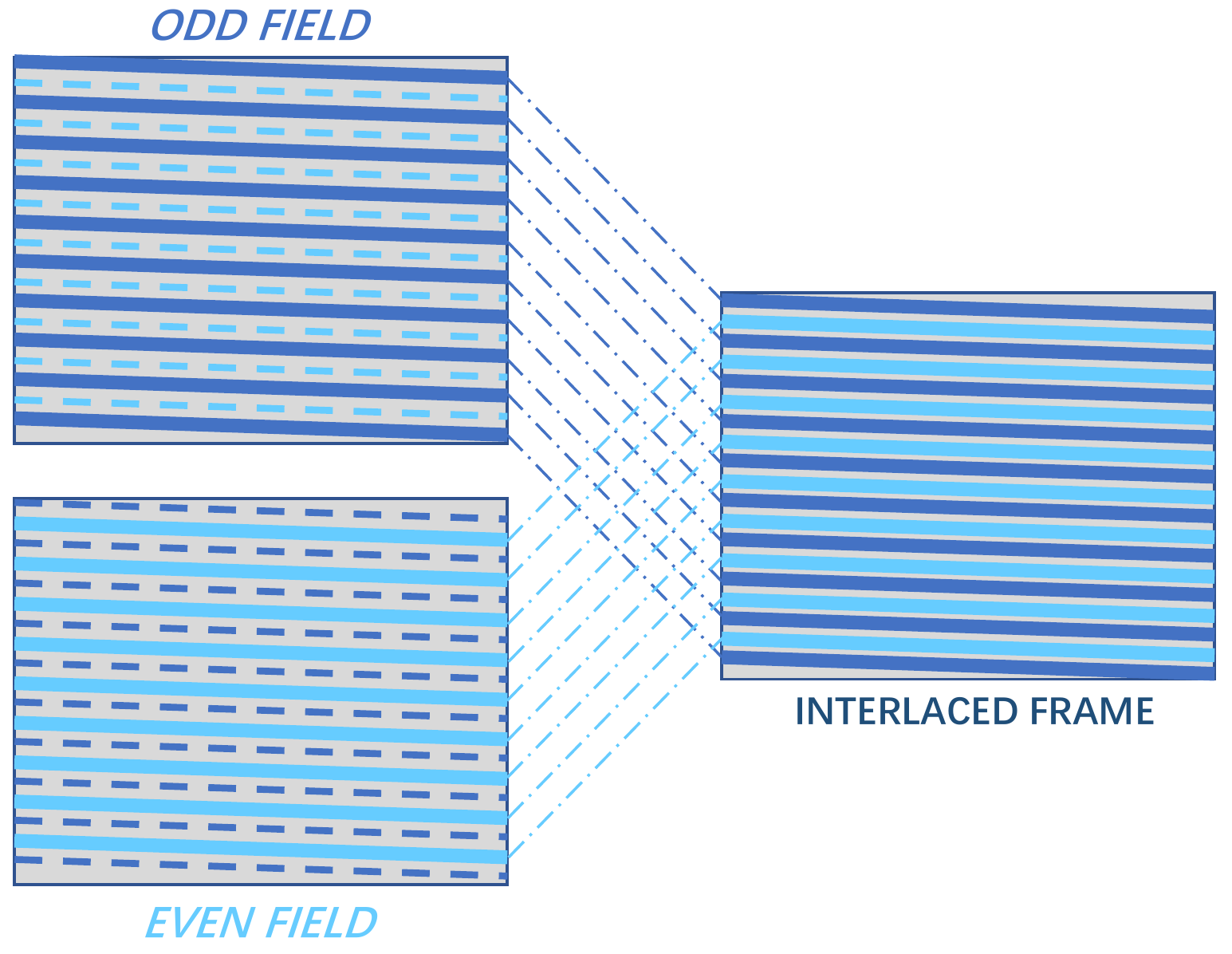}
	\caption{Illustration of the interlaced scanning mechanism.}
	\label{fig:1}
\end{figure}

\begin{figure}[htb]
	\setlength{\abovecaptionskip}{0.2cm}
	\setlength{\belowcaptionskip}{0.2cm}
	\centering
	\includegraphics[height=5cm,width=0.4\textwidth]{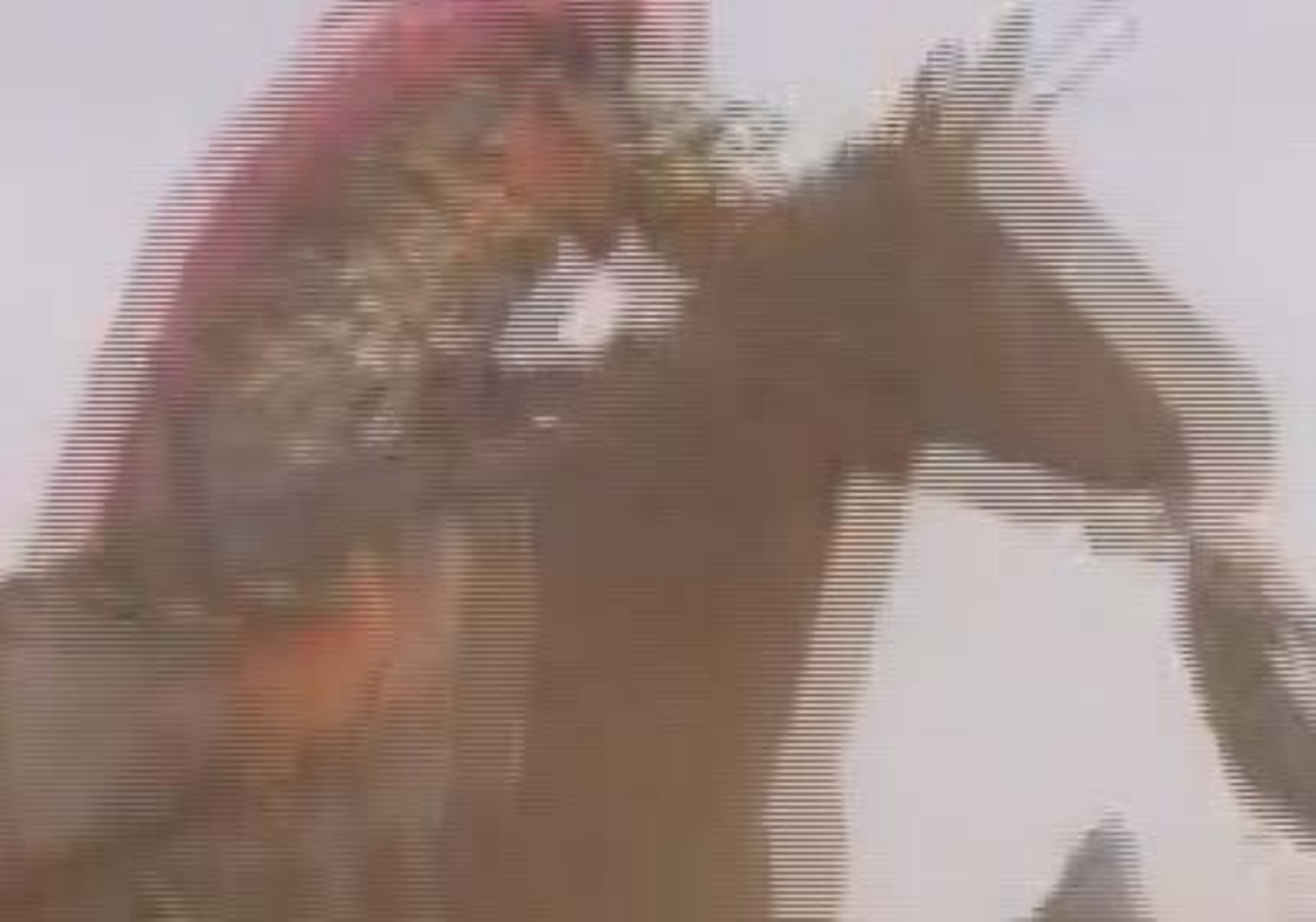}
	\caption{Interlaced frame may contain severe comb-teeth artifacts.}
	\label{fig:2}
\end{figure}

The traditional interlaced scanning mechanism can be defined as $I=f_{interlace}(T_{odd},T_{even})$, where $I$ is an interlaced frame,$T_{odd}$ and $T_{even}$ denote the odd field and even field respectively, and $f_{interlace}$ represents the interlacing scanning process as shown in Fig. 1. Previous deinterlacing methods mainly solve the inverse process of $f_{interlace}$. Among them, the traditional methods can be categorized into temporal interpolation, spatial interpolation and motion-based methods. Temporal interpolation-based method utilizes adjacent frames to interpolate the lost lines, but it is sensitive to large motion and may produce severe motion artifacts. Spatial interpolation-based method uses only one field to estimate the whole frame to avoid interlacing aliasing brought by large motions. However, due to the missing information of another field, the quality of reconstructed details is significantly reduced. Motion-based method first estimates the motion between two frames and then performs temporal or spatial interpolation based on the estimated motion. With the rapid development of deep neural network (DNN) and video processing techniques, many effective DNN-based image and video restoration models \cite {zhang2017beyond, dong2015image, lim2017enhanced} have been proposed. But they cannot solve deinterlacing task directly, since interlacing is caused by specific manmade mechanisms. DICNN \cite{zhu2017real} first introduced shallow convolutional networks to the real-time video de-interlacing task. However, this model still ignores the mixture of various kinds of unnatural effects in real-world early interlaced videos. Besides, the shallow backbone of DICNN leads to limited learning ability, and thus cannot handle the complex artifacts in early videos well. 

In this paper, the degradation model of early interlaced videos is defined as $I=f_{codec}(f_{interlace}(T_{odd},T_{even} ))+n$, where $f_{codec}$ denotes the video compression process and $n$ represents noises. Existing jointly deinterlacing and denoise methods \cite{zhao2020din,2020Deep} are based on single-frame reconstruction and cannot make full use of the temporal similarity between adjacent frames. However, interlaced frames are highly correlated with adjacent frames in time series, hence the performance of current single-frame de-interlacing networks \cite{zhu2017real,zhao2020din,2020Deep} still has a lot of room for improvement.}

In order to solve the above problems, we propose a multi-frame early video deinterlacing network (MFDIN), which aggregates information of multiple frames for reconstruction.  The MFDIN is composed of three modules: , i.e., spatial feature interpolation module, temporal alignment fusion module and frame reconstruction module. To make fully use of the prior knowledge of the interlacing scanning mechanism, the input frames are firstly split into odd and even fields and then input to the first module for feature extraction and vertical pixel interpolation.Inspired by recent state-of-the-art (SOTA) super-resolution algorithm \cite{wang2019edvr} and deformable convolution (DCN) methods \cite{dai2017deformable, zhu2019deformable}, an efficient alignment fusion module was proposed to effectively perform temporal alignment and aggregation by learning more applicable offsets, which can reduce the negative impact of complex artifacts such as compression artifacts on inter-frame motion. {\color{cblue}In addition, the MFDIN can support the output of same frame rate or double frame rate, because each interlaced frame naturally contains two fields.}

The main contributions of this paper are summarized as follows:{\color{cblue}
\begin{enumerate}
\item This paper focuses on joint enhancement of early interlaced videos in real world, rather than traditional deinterlacing. An effective multi-frame deinterlacing network is proposed to fully use the temporal information, while current early video deinterlacing methods only focus on single frame.

\item Appropriate field vertical interpolation and lightweight alignment fusion module (AFM) have been specifically designed for this complex task. An efficient offset network is presented to perceive interframe motion accurately. Comparing to multiscale DCNs, the proposed AFM with single DCN can improve the stability and reduce the computational cost simultaneously. 

\item Owing to the specifically designed multi-frame architecture, MFDIN can achieve convincing and visually appealing results on both synthetic datasets and real-world early interlaced videos.

\item Furthermore, extended experiments on different tasks, e.g., 1080i(25P)-4K(50P), demonstrate that the proposed model is suitable for the joint problem of deinterlacing, compression artifacts removal, video super-resolution and double frame rate.
\end{enumerate}
}
The rest of the paper is organized as follows. Related works are reviewed in Section II. Section III introduces the proposed method in detail. Experimental results are discussed in Section IV, and Section V concludes the paper.

\begin{figure*}[htb]
	\setlength{\abovecaptionskip}{0.2cm}
	\setlength{\belowcaptionskip}{0.2cm}
	\centering
	\includegraphics[height=6.2cm,width=0.9\textwidth]{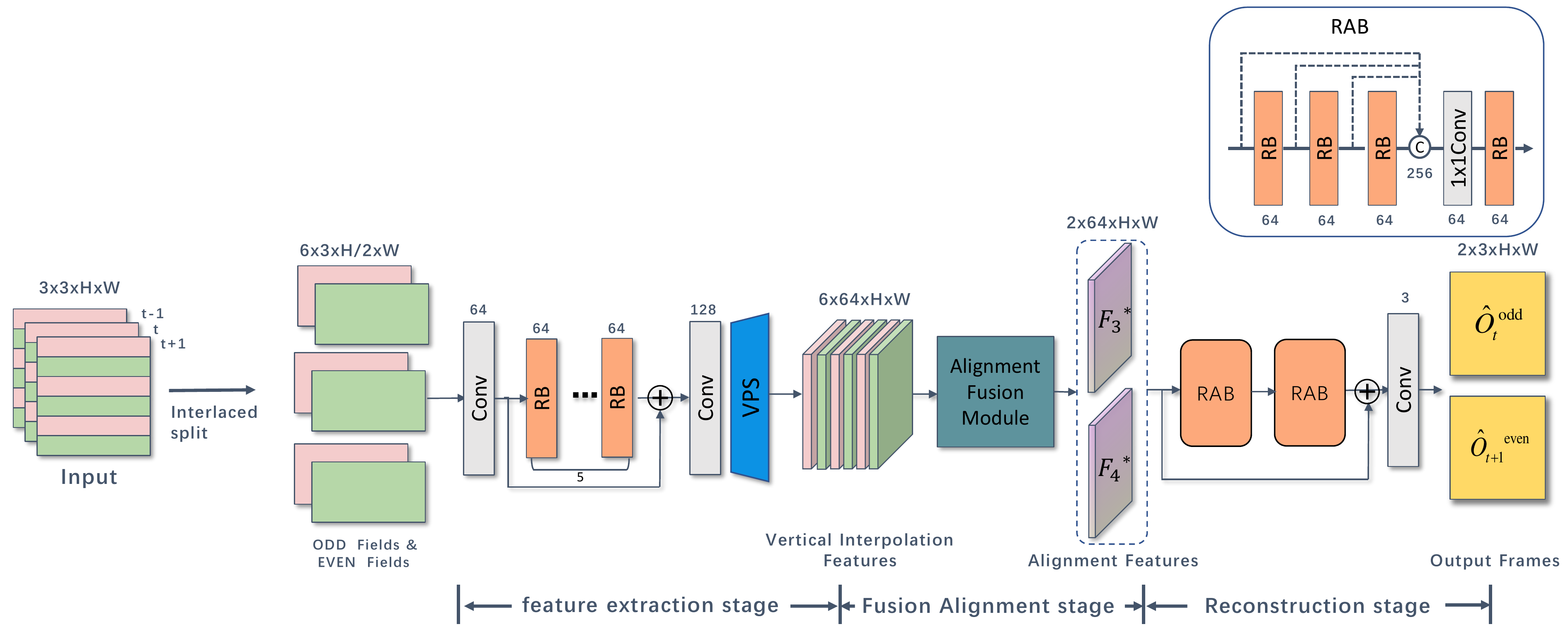}
	\caption{{\color{cblue}Architecture of the proposed multi-frame deinterlacing network (MFDIN). Three input frames and two output frames are used as an illustrative example.}}
	\label{fig:3}
\end{figure*}
\section{Related works}
\subsection{Video Deinterlacing.}
Traditional deinterlacing methods \cite{de1998deinterlacing, doyle1998interlaced} mainly focus on estimating the missing lines in each field, so as to realize the effect of progressive scanning. Related algorithms can be roughly divided into temporal interpolation, spatial interpolation, and motion-based methods. Temporal interpolation algorithm is a basic way of deinterlacing, which estimates missing lines by means of adjacent frames. However, this type of method may produce severe artifacts when there is a large motion. Spatial interpolation methods \cite{wang2012efficient, wang2013moving, jakhetiya2014fast, brox2014edge} estimate the missing lines via spatial interpolation of a single field. Because the interlacing artifacts tend to appear at the edges of objects, these methods focus on reducing the step effect at the image boundary. However, because only half a frame is used, the results are usually blur and less than satisfactory. To reproduce better results, motion-based methods \cite{mohammadi2012enhanced, jeon2009weighted, zhu2017motion} combine multiple fields in adjacent frames based on accurate motion adaptation or motion compensation. However, accurate motion is difficult to obtain and the computational complexity of these methods is usually too high. In recent years, DNN-based methods \cite{sajjadi2018frame, li2019feedback, zhang2018adversarial, zamir2017feedback, xiang2020zooming} have achieved much better performance than traditional methods in many low-level vision tasks. Zhu \textit{et al.} \cite{zhu2017real} first introduced a shallow deinterlacing convolutional neural network (DICNN) to video deinterlacing tasks and achieved impressive results. However, the learning ability of shallow networks is limited, and complex artifacts in real-world interlaced videos cannot be handled well in this manner. Zhao \textit{et al.} \cite{zhao2020din} proposed a deeper deinterlacing network (DIN) for early videos by vertically up-sampling and merging two fields. As mentioned before, these DNN-based methods are all based on a single frame, which cannot sufficiently utilize inter-frame redundancy information. To improve deinterlacing performance, we first propose a multi-frame-based method, which aggregates information from multiple fields through lightweight and efficient alignment, to restore high visual quality frames.


\subsection{Related Multi-Frame Video Restoration.}
{\color{cblue}Compared to single-frame restoration \cite{lim2017enhanced, zhang2018density, yu2018crafting, yu2019path},  video restoration tasks, such as video super resolution (VSR) and video compression artifact removal, usually utilize more temporal information to refine missing details.} But the ubiquitous occlusion and large motion in the frame sequence seriously affect the performance of the networks. Therefore, inter-frame information often requires temporal alignment or other processing to be fully utilized. The basic approach is to estimate the motion between frames by calculating the optical flow \cite{sun2018pwc, xue2019video, caballero2017real}. However, for the areas contain occlusion and large motion, it is difficult and high-computational-cost to obtain accurate flow. 3D convolution can also be used to achieve temporal aggregation \cite{kim20183dsrnet,jo2018deep,ying2020deformable}. For instance, Joe \textit{et al.} \cite{jo2018deep} used stacked 3D convolution layers for motion compensation to extract temporal information. Ying et al. \cite{ying2020deformable} adopted 3D deformable convolution to extract spatial and temporal information within a temporal sliding window for VSR and implemented this method over the entire video sequence. However, these methods lead to massive redundant computation, which limits the efficiency of the model. {\color{cblue}Some methods replace the above scheme by implicit alignment. For instance, TDAN \cite{tian2020tdan} and EDVR \cite{wang2019edvr} reconstructed high quality super-resolution frames through DCN-based alignment modules, which surpasses some flow-based methods. Niklaus et al. \cite{niklaus2017video} and zhou et al. \cite{zhou2019spatio} proposed adaptive convolution for video interpolation and video deblurring by implicitly utilizing motion cues and pixel information. In addition, recurrent neural networks (RNN) have also been introduced to video tasks to simplify sequence-to-sequence learning. For instance, a modified convolutional long short-term memory Network (LSTM) was introduced for video quality enhancement in \cite{yang2019quality}.  Isobe et al. \cite{isobe2020revisiting} proposed a recursive time modeling method with residuals to stabilize the training of RNNs and improve the performance. In \cite{guan2019mfqe}, Bi-LSTM was used to distinguish high quality  frames in the compressed sequence, so as to help enhance the reconstruction of the adjacent low-quality compressed frames.} For our task, the mixed interlacing and other artifacts increase the difficulty of explicit motion estimation. Inspired by \cite{wang2019edvr}, we propose an efficient alignment module, which implicitly aligns and fuses features from multiple fields by means of more accurate offset estimation.

\section{Proposed Method}
\subsection{Overview}
Given 3 consecutive interlaced frames $I_{[t-1:t+1]}$, our purpose of deinterlacing is to reconstruct a high-quality central frame $\hat{O}_{t}^{odd}$, which is consistent with the ground truth frame $O_{t}^{odd}$. Note that for the tasks that need to double the frame rate at the same time, \textit{e.g.} 25P-to-50P, the proposed model can also reconstruct two consecutive frames $\hat{O}_{t}^{odd}$and $\hat{O}_{t}^{even}$ for each central frame $I_t$. As illustrated in Fig. 3, the proposed MFDIN consists of three stages, \textit{i.e.}, feature extraction, alignment fusion, and reconstruction. Firstly, we divide each frame into odd and even fields for feature extraction and vertical magnification. In order to align and aggregate the features of multiple fields, the reconstructed features of adjacent frames and central frame are then gathered by the alignment fusion module. Finally, the aligned and fused multi-frame features pass through the reconstruction module, to further reduce artifacts and restore high-quality results. Each module will be introduced and analyzed in detail in the following.

\begin{figure}[htb]
	\setlength{\abovecaptionskip}{0.2cm}
	\setlength{\belowcaptionskip}{0.2cm}
	\centering
	\includegraphics[height=3cm,width=0.36\textwidth]{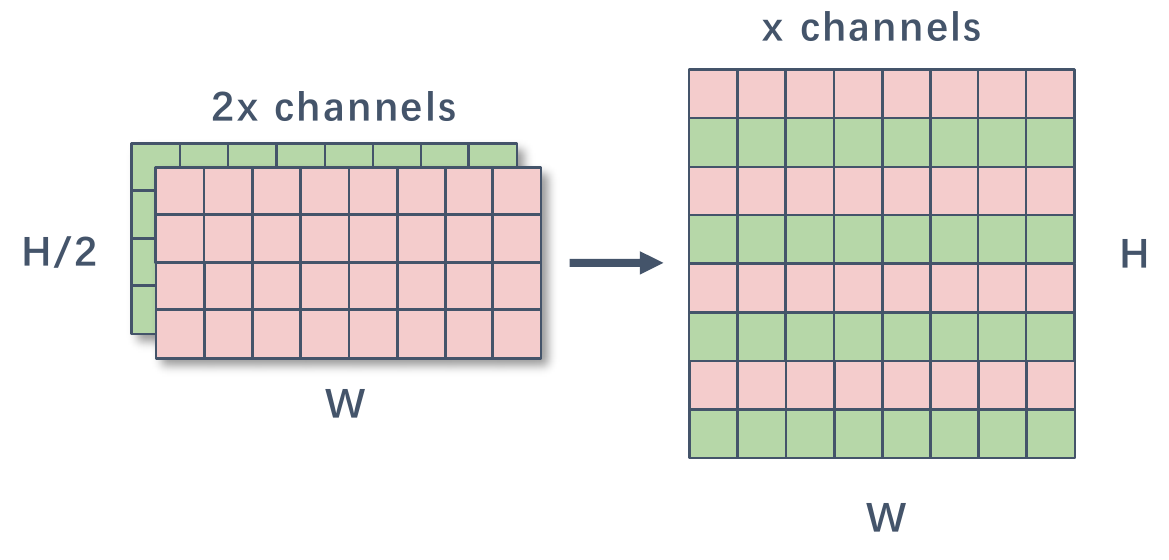}
	\caption{Illustration of the vertical pixel shuffle layer.}
	\label{fig:4}
\end{figure}
\begin{figure}[htb]
	\setlength{\abovecaptionskip}{0.2cm}
	\setlength{\belowcaptionskip}{0.0cm}
	\centering
	\includegraphics[height=4.4cm,width=0.48\textwidth]{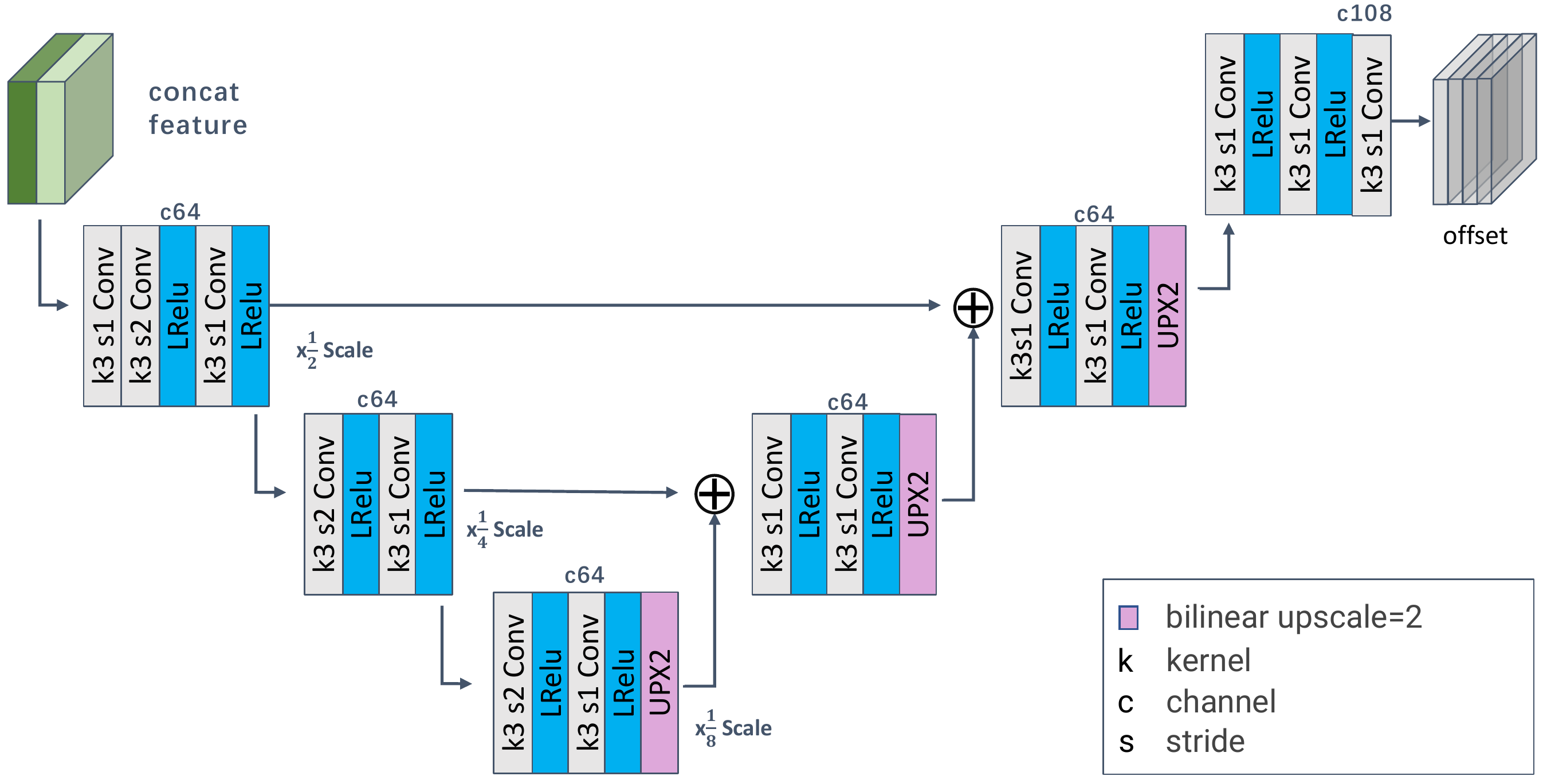}
	\caption{Architecture of the offset net.}
	\label{fig:5}
\end{figure}
\begin{figure}[H]
	\setlength{\abovecaptionskip}{0.2cm}
	\setlength{\belowcaptionskip}{0.2cm}
	\centering
	\includegraphics[height=4.2cm,width=0.48\textwidth]{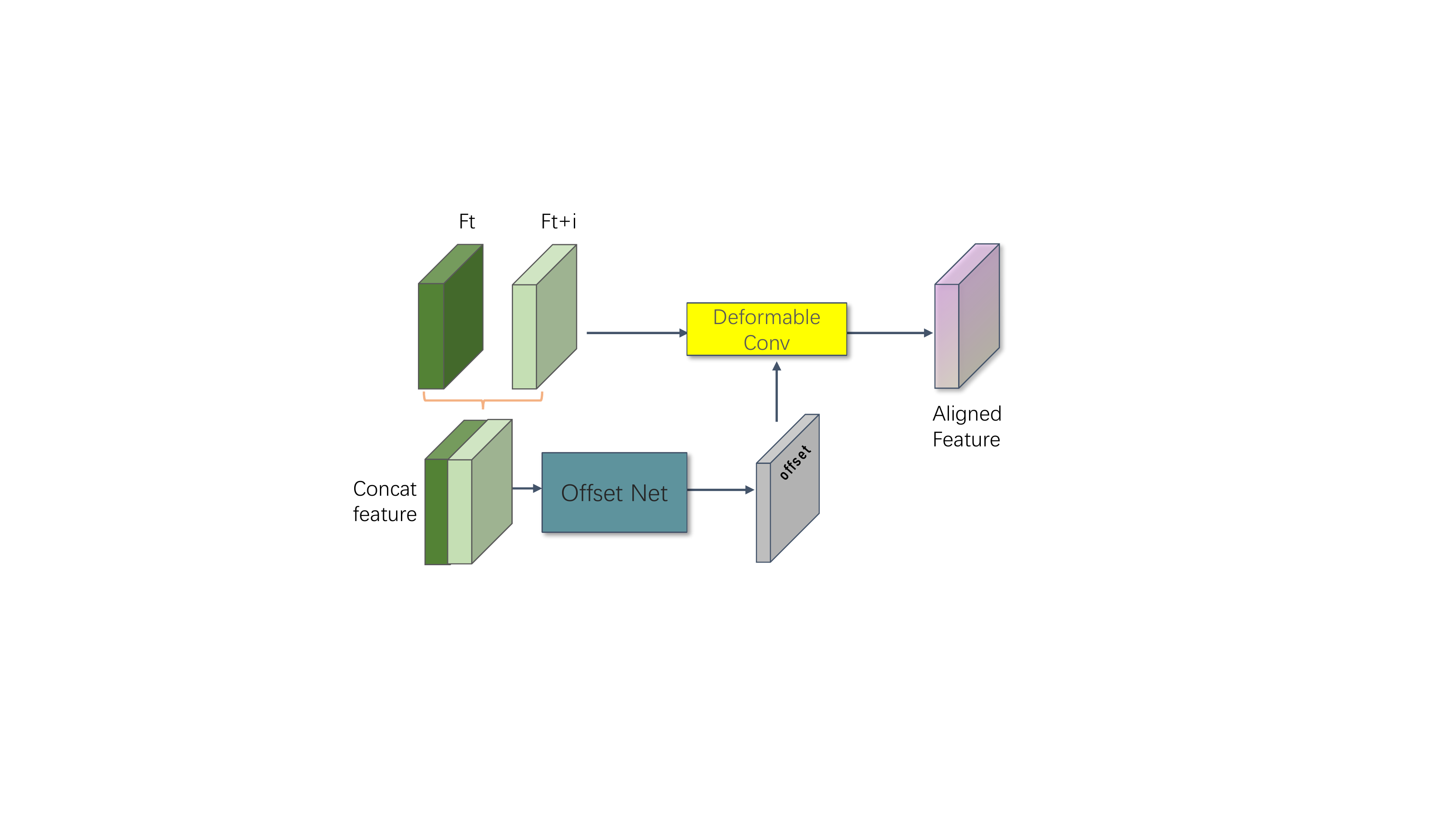}
	\caption{Alignment fusion module.}
	\label{fig:6}
\end{figure}

\subsection{Feature extraction module}
The main purpose of the feature extraction module is to extract features from split fields, reproduce vertically interpolated frames, and reduce the combined artifacts preliminarily.
The classical ResNet \cite{he2016deep}, which is commonly used in related low-level vision tasks \cite{lim2017enhanced} is selected as the backbone. Motivated by DIN\cite{zhao2020din}, the odd and even fields of each frame are separately input for feature extraction, which can simulate the reverse process of manmade interlacing scanning mechanism. In this way, the network can perceive the difference among multiple fields, and further reduce the impact of interlacing artifacts during feature extraction. Note that the resolution of the input fields is half of the original frame, we interpolate the reconstructed field features in the vertical direction to restore the original resolution by means of a vertical pixel shuffle  (VPS) layer \cite{shi2016real,zhao2020din}, which is illustrated in Fig. 4. Specifically, as shown in Fig. 3, the three input interlaced frames are firstly divided into six fields, and then these fields are separately input into the same feature extraction module to obtain six reconstructed frames with the original resolution. In this paper, the feature extraction module contains five residual blocks (RB) \cite{lim2017enhanced} and global skip connection is also utilized to improve the convergence of the algorithm.

After the feature extraction stage, interlacing artifacts in reconstructed frame features can be effectively reduced. However, many high frequency details and complex noises cannot be handled well because only single-field information is used. Hence, temporal information can be aggregated to obtain better reconstruct results.

\subsection{Alignment fusion module}
In multi-frame schemes, the neighbor frames need to be accurately aligned to avoid ghost artifacts. For deinterlacing task, common optical flow alignment or 3D convolution \cite{tran2015learning} methods may be suboptimal. Due to a variety of complex artifacts, it is difficult to learn accurate optical flow, and inaccurate optical flow seriously affects the performance. In addition, gradually fusing spatial-temporal information through multiple stacked 3D convolution layers creates a huge computational burden, which reduced the overall efficiency of the model. In this paper, we tend to find an efficient way to fuse temporal information with low computational complexity.

Inspired by SOTA multi-frame super-resolution method EDVR \cite{wang2019edvr}, we also use DCN-based implicit alignment instead of explicit optical-flow-based warping. In EDVR \cite{wang2019edvr}, multiple DCNs are used at multiple scales to enhance performance, but this scheme is not very suitable to our task. Firstly, although stacking multiple DCNs can enlarge the receptive field of DCN, the network fluctuates frequently in the training process, which often results in offset explosion in our de-interlacing experiment, making the network difficult to converge. Moreover, the original DCN structure is used directly for most video tasks, but the offset required by the original deformable convolution for target detection is different from that for temporal alignment. The motion flow representation we used has always been considered that is intuitively and empirically invariant with appearance. Especially in our task, the feature information contains a large number of complex artifacts, which further increases the difficulty of offset learning. Therefore, traditional DCN, which merely learns the offset with a 3$\times$3 convolution, should not be simply adopted. Hence, we design a lightweight offset network, which follows the multi-scale philosophy that commonly used in optical flow estimation \cite{dosovitskiy2015flownet}. The U-net \cite{ronneberger2015u} structure with large receptive fields and multiple scales is adopted to dynamically deal with different displacements, and to predict more appropriate offsets. Detailed architecture of the offset network is shown in Fig. 5. Different from the common optical flow alignment model, implicit alignment is directly performed on the feature level from the reconstructed multiple frame sequences, as illustrated in Fig. 6. 

Given the six feature maps $F_{i} (i=1,2,3,4,5,6)$ reconstructed in the previous stage, our goal is to obtain aligned and fused features $F_{3}*$ and $F_{4}*$ of the central two fields. Taking $F_{3}^*$ as an example, each adjacent map is concatenated with $F_{3}$ and then corresponding offset $\Delta P$ is calculated as:
\begin{figure}[ht]
	\centering
	\includegraphics[height=4cm,width=0.35\textwidth]{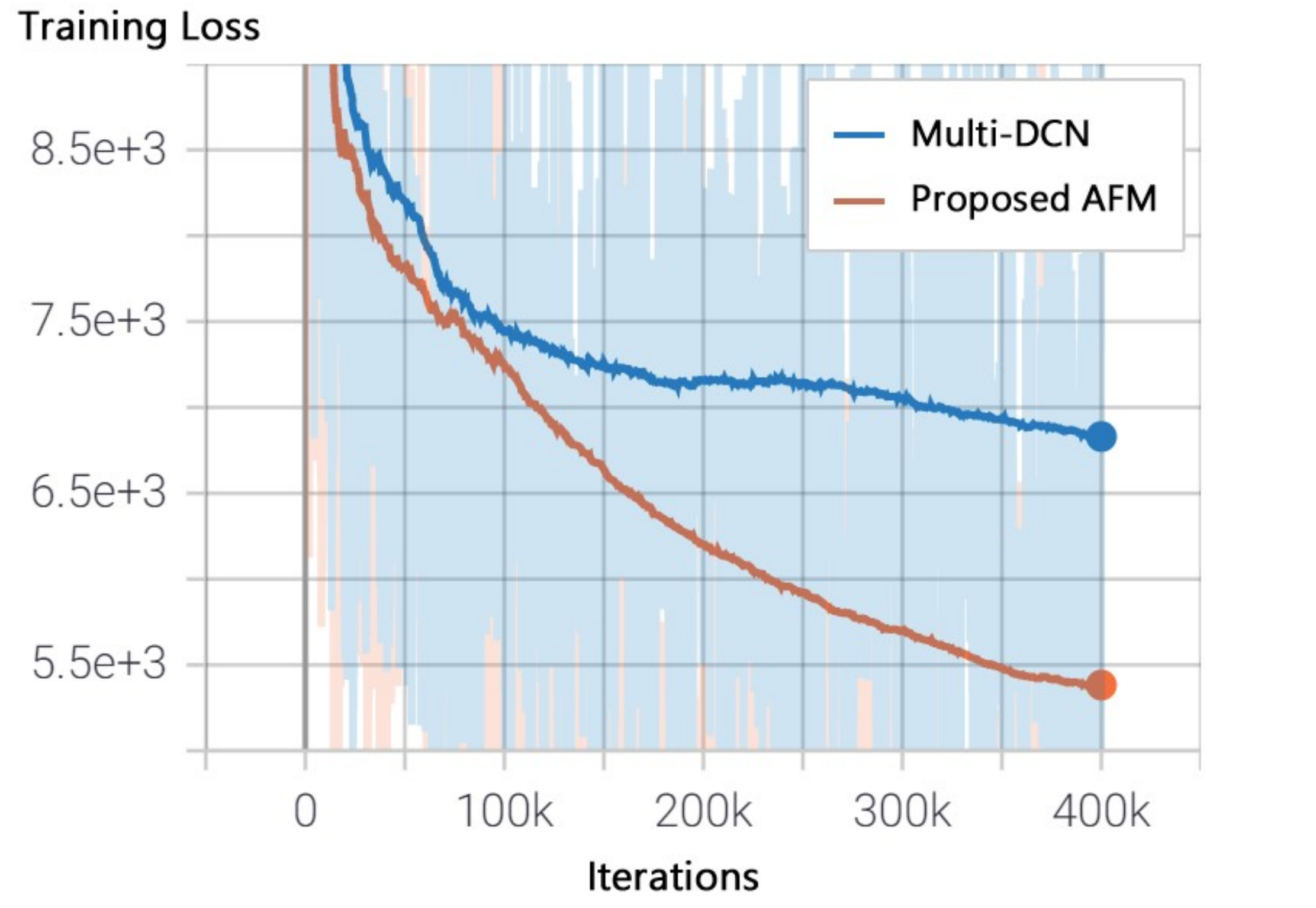}
	\caption{{\color{cblue}Comparison of the loss curves of multi-DCN scheme and the proposed alignment fusion module.}}
	\label{fig:7}
\end{figure}
\begin{equation}\label{eqn2}
	 \Delta P_{j}= f([F_{3}, F_{j}]),  j=(1,2,4,5)
\end{equation}
{\color{cblue}where $f(\cdot)$ denotes the offset net, and [·,·] denotes the concatenation operation. Here, $\Delta P_j = \left\{\Delta p_{k} | k=1,2,...,K\right\}$ refer to the offsets of the K sampling positions of the convolution kernel. For instance, 9 sampling positions are corresponding to each $3 \times 3$ convolution kernel. With the learned offset $\Delta P_j$, dynamic sampling can be performed by means of DCN to achieve feature alignment:
\begin{equation}\label{eqn3}
	A_{j}= f_{DCN}(F_{j},\Delta P_{j}),  j=(1,2,4,5)
\end{equation}
where $f_{DCN}(\cdot)$ is the deformable convolution operation. More specifically, for each position $p_0$ on the aligned feature $A_{j}$, the dynamic sampling process can be defined as:
\begin{equation}\label{eqn1}
	A_{j}(p_{0})=\sum_{k=1}^{K}w_{k}\cdot F_{j}(p_{0}+\Delta p_{k})
\end{equation}
where $w_k$ is the kernel weight, and the convolution will be operated on the irregular positions $p_{0}+\Delta p_{k}$. Then, the fusion features can be obtained:}
\begin{equation}\label{eqn4}
	F_{3}* = g([A_{j},]) ,  j=(1,2,3,4,5)
\end{equation}
where $g(\cdot)$ denotes a $1\times1$ convolutional layer that is used to fuse aligned features. Similarly, $F_{4}^*$ is calculated in the same way as $F_{3}^*$ with the corresponding neighbor features $F_j (j=2,3,5,6)$.

{\color{cblue}Compared with the multiple-DCN scheme, the proposed alignment fusion module can estimate more applicable offsets, thus making the network more stable in the training phase. Fig. 7. shows the loss curves of these two schemes. It can be found that the proposed AFM performs better in terms of both training stability and convergence speed, which proves that the designed offset network structure can learn more accurate offsets and thus achieve better aligned and fused features in the case of complex degradations.}

\subsection{Frame Reconstruction module}
To further restore the details and refine the visual quality, the fused features from the previous stage are input into the reconstruction module, which is composed of 8 stacked RBs.
Motivated by the feature aggregation structure \cite{liu2020residual}, we introduce two residual aggregate blocks (RAB) to utilize the features in different levels. The structure of RAB is shown in Fig. 3, which concatenates the outputs of multiple RBs and then fuses them via a $1\times1$ convolutional layer. Finally, the features are reconstructed to one or two output frames by means of an output layer, which is a $3\times3$ convolutional layer.
To optimize our network, Charbonnier penalty function \cite{lai2017deep} is adopted:
\begin{equation}\label{eqn4}
	Lr=\sqrt{||\hat{O}_{t}^{odd}-{O}_{t}^{odd}||^{2}+\varepsilon^{2}}	
\end{equation}
When two frames are restored from one interlaced frame at the same time, the loss function becomes:
\begin{equation}\label{eqn4}
	Lr=\sqrt{||\hat{O}_{t}^{odd}-{O}_{t}^{odd}||^{2}+||\hat{O}_{t}^{even}-{O}_{t}^{even}||^{2}+\varepsilon^{2}}
\end{equation}
where $\hat{O}_{t}^{odd}$ and $\hat{O}_{t}^{even}$ refer to the output frames of the network, ${O}_{t}^{odd}$ and ${O}_{t}^{even}$ refer to ground-truth HR video frame, and $\varepsilon$ is empirically set to $1\times10^{-3}$.

\section{EXPERIMENTS}
\subsection{Datasets}
Because high-quality ground truth of early interlaced video is difficult to be obtained, we build a synthetic training dataset based on YOUKU 2K video dataset \cite{YoukuVSRE2019}, and constructed three test sets to conduct objective quality assessment. Besides the synthetic datasets, several real-world early PAL/NTSC TV videos are used for comparison and subjective tests. In addition, extended experiments in different application scenarios are also implemented, such as 540i-to-1080p, and 1080i(25P)-to-4K(50P).

\textbf{Training Set.} {\color{cblue}YOUKU-2K \cite{YoukuVSRE2019} is an open-source video super-resolution dataset, which is composed of 1000 2K video sequences in RAW format. Note that this dataset contains various types of films and TV series, which are highly consistent with the contents of early interlaced TV videos.} For training, we collected 400 video sequences (excluding cartoons) from the dataset and then generated interlaced frames according to the traditional interlaced scanning mechanism, \textit{i.e.}, odd and even lines of an interlaced input frame are scanned from two adjacent ground truth frames. Note that two ground truth frames are thus corresponding to one interlaced frame. For the task that only restores one frame for an interlaced frame, the first ground truth frame is selected. Furthermore, to roughly simulate the mixed compression artifacts in early videos, we applied the FFMPEG tool to compress the synthesized interlaced videos with H.264 codec with random "-CRF" parameters from 30 to 38.

\textbf{Synthetic Testing Sets.} For testing, we conduct three synthetic testing sets. First, we selected 7 videos (70 groups of frames) from the remaining YOUKU-2K videos. Then, 40 groups of frames were also selected from SJTU-4K dataset \cite{song2013sjtu} to verify the reconstructed results for ultra-high-definition videos. {\color{cblue}Furthermore, 150 groups of frames were selected from the Vimeo90K dataset \cite{2019Video}, which is widely used in the video enhancement tasks. }The frames in these testing sets were degraded in the same way as the training set.

\begin{figure*}[h]
	\centering
	\includegraphics[height=6.2cm,width=\textwidth]{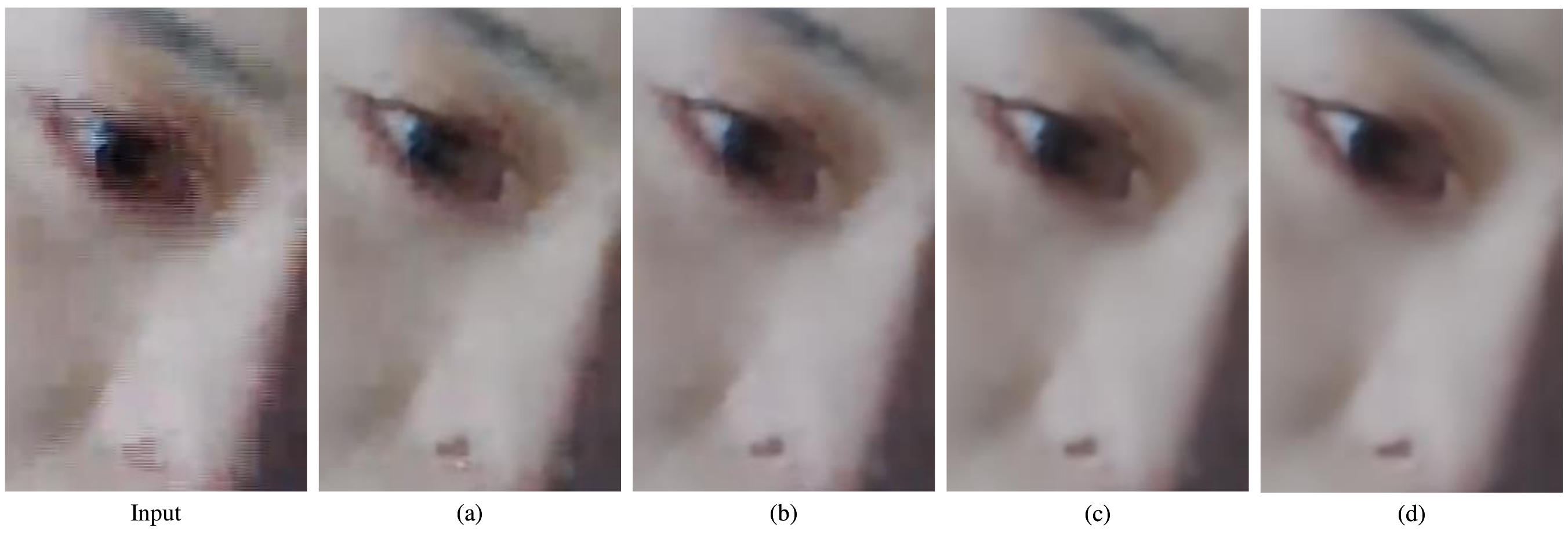}
	\caption{{\color{cblue}Results of different methods in ablation testing, (a)model with single frame as input, (b)model with multi frames as input, (c)model with split frame, (d)model with split frame and alignment.}}
	\label{fig:8}
\end{figure*}
\begin{table}[htb]
	\renewcommand{\arraystretch}{1.5}
	\caption{{\color{cblue}Psnr Results of Ablation Testing on Synthetic Datasets}}
	\label{table:1}
	
	\centering
	\tiny
	\begin{tabular}{ccccc}
		\hline
		\textbf{Model}              & \textbf{Multi Frames} & \textbf{Split Frame} & \textbf{Alignment Fusion} & \textbf{PSNR}  \\ \hline
		Baseline                    &                      &                       &                           & 29.50          \\
		Multi Frames(entire)         & $\surd$                    &                       &                           & 31.44          \\ \hline
		MFDIN(one frame with split) &                      & $\surd$                     & $\surd$                         & 32.19          \\
		MFDIN(without split frame)  & $\surd$                    &                       & $\surd$                         & 32.32          \\
		MFDIN(without aligment)     & $\surd$                    & $\surd$                     &                           & 32.18          \\
		MFDIN                       & $\surd$                    & $\surd$                     & $\surd$                         & \textbf{32.76} \\ \hline
	\end{tabular}
	
\end{table}

\begin{table}[htb]
	\renewcommand{\arraystretch}{1.5}
	\caption{{\color{cblue}Psnr Results of MFDIN with Different number of input frames}}
	\label{table:2}
	\centering
	\begin{tabular}{ccc}
		\hline
		Number of input frames           & PSNR  & FLOPS \\ \hline
		1 frame(2fields) & 32.19 & \textbf{550G}  \\
		3 frames(5fields) & \underline{32.76} & \underline{1375G} \\
		5 frames(7fields) & \textbf{32.88} & 1834G \\ \hline
	\end{tabular}
	
\end{table}

\textbf{Real-world Early Video Testing Set.} A total of 8 PAL/NTSC TV videos, which contain severely interlaced artifacts, were collected to verify the generalization of the proposed method on real-world early videos. Note that there lacks ground truth for these real-world videos, hence, subjective evaluation is implemented by comparing the visual quality.

\textbf{Extended Training and Testing Sets.} In addition to the complex artifacts, early videos are often with low resolution and low frame rate. To verify the proposed method in different application scenarios that contain deinterlacing, compression artifacts removal, super-resolution and reproduce double frame rate, we conducted extended experiments on different application scenarios, \textit{i.e.}, 1080i-to-4K, 1080i(25P)-to-4K(50P), 540i-to-1080p, and 540i(25P)-to-1080p(50P). To build training and testing sets in these cases, we first used FFMPEG tool to perform $2\times$ down-sampling and frame rate adjustment, and then degraded these videos via the aforementioned interlacing and compression process.

\subsection{Implementational details}
In our experiment, we selected more than 28,000 groups of frames from the training set, and each group consists of three consecutive synthetic interlaced frames. These frames were further cropped at random position with the resolution of $128\times128$ in the training process. The number of channels in each RB is set to 64. Moreover, in the alignment fusion module, the number of channels in the offset network is set to 64 at all scales. For optimization, we use Adam optimizer \cite{kingma2014adam}, the learning rate is initially set as $4\times10^{-4}$, and then decayed to $1\times10^{-7}$ with a cosine annealing. The proposed MFDIN is implemented with the PyTorch framework and trained 100 epochs on a NVIDIA TATAN V GPU.

\subsection{Ablation studies}
Ablation testing is conducted to verify the effectiveness of different components in MFDIN. Firstly, we compare the reconstruction results of single frame input and multi-frame input. By comparing the results in Fig. 8(a) and (b), multi-frame method can restore better details than the single-frame method due to the more information is available from adjacent frames. By comparing Fig. 8(b) and (c), we can find that input split fields, which utilizes the prior knowledge of manmade interlaced scanning, can reproduce better results than input whole frames. As shown in Fig. 8(d), the proposed alignment module can effectively improve the utilization of temporal information and reduce the negative impact of large inter-frame motion on the reconstruction result, so as to restore clearer images. {\color{cblue}At the same time, the quality assessment results are listed in Table I. Compared to the complete model, each of the proposed strategies brings different degrees of improvement. Even only one frame is input, the proposed module can provide significant gain after applying field splitting and alignment. Because one interlaced frame is equivalent to two consecutive fields, and the effective alignment further improves the utilization and fusion of information.}

{\color{cblue}In addition, Table II lists the PSNR and FLOPS of the reconstructed results with different input frames, i.e., 2 fields (1 frame), 5 fields (3 frames) and 7 fields (5 frames). It can be found that multi-frame methods can achieve better performance by using more adjacent frames. Note that, although adding more neighbor frames can further refine the PSNR results, but it also significantly increases the computational cost. In our experiment, 5 neighbor fields (from 3 adjacent frames) are chosen as input to balance performance and cost.}

\begin{figure*}[htp]
	\centering
	\includegraphics[height=10cm,width=\textwidth]{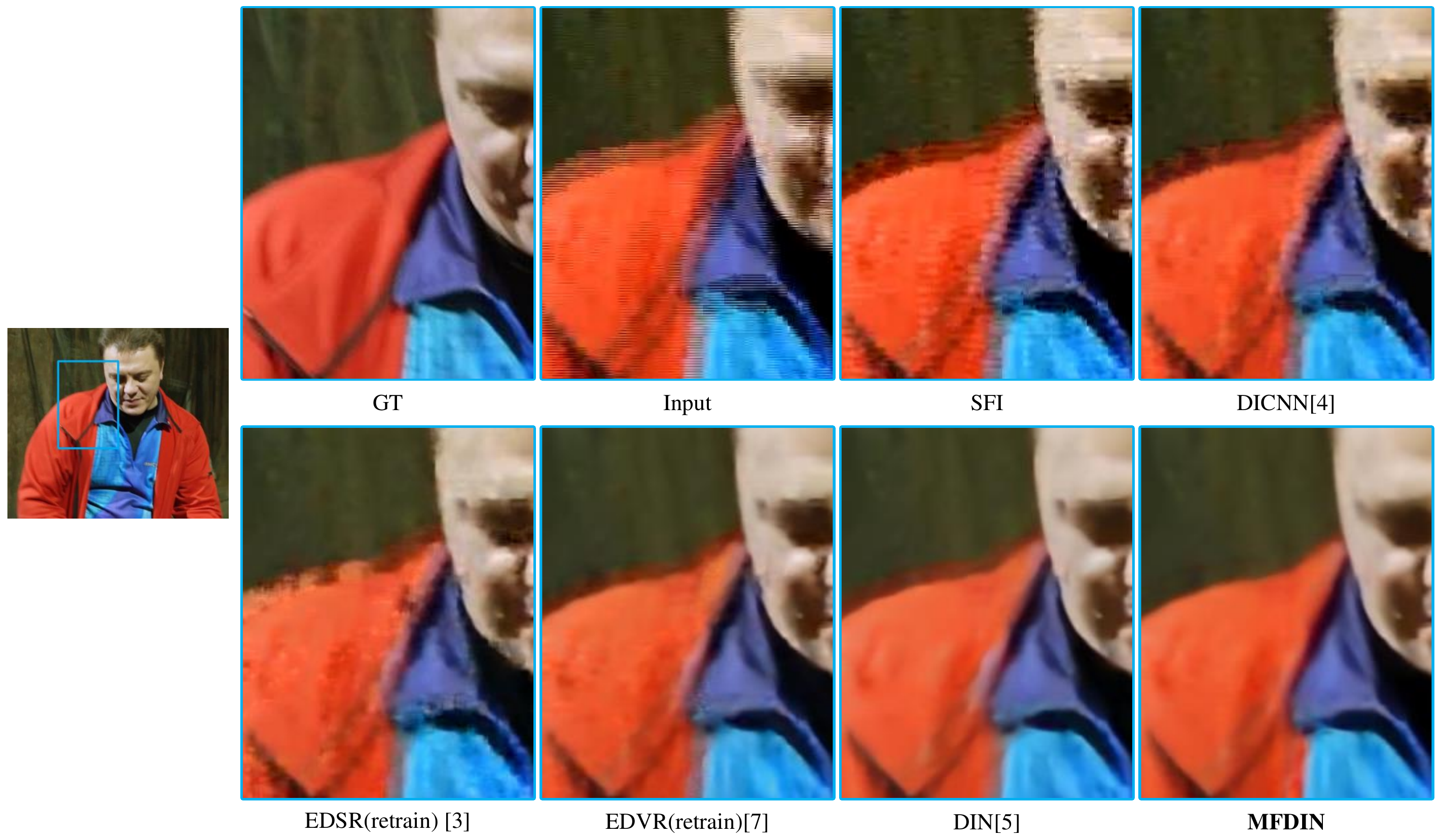}
	\caption{Qualitative comparison on the \textbf{YOUKU 2K} synthetic testing set.}
	\label{fig:9}
\end{figure*}

\begin{figure*}[htp]
	\setlength{\abovecaptionskip}{0.2cm}
	\setlength{\belowcaptionskip}{0.0cm}
	\centering
	\includegraphics[height=5.6cm,width=\textwidth]{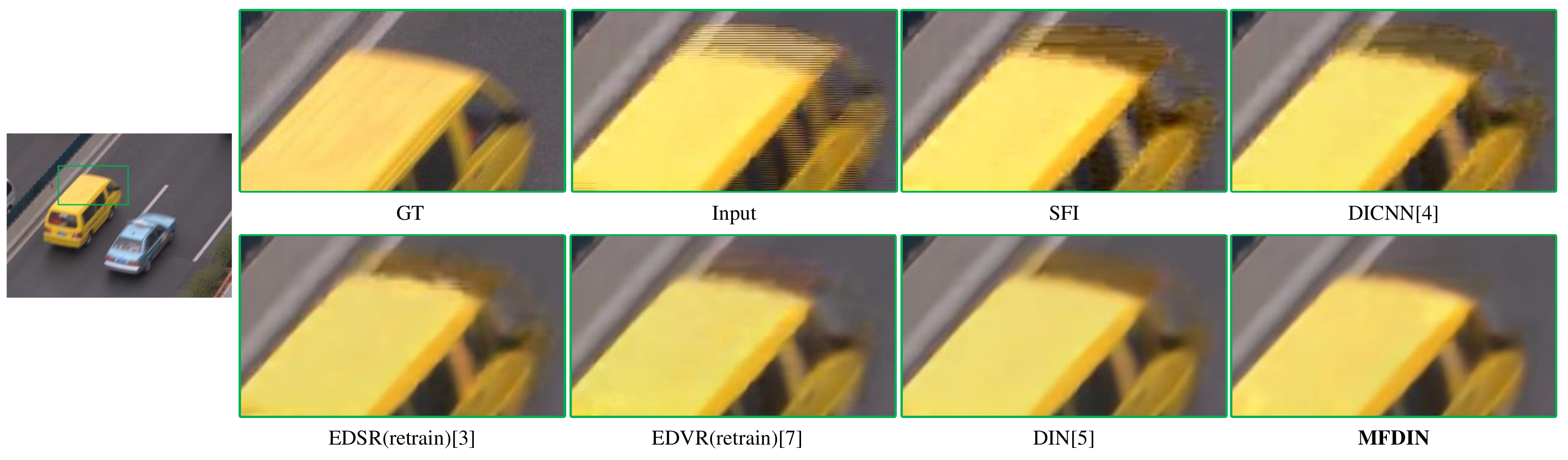}
	\caption{Qualitative comparison on the \textbf{SJTU 4K} synthetic testing set.}
	\label{fig:10}
\end{figure*}
\begin{figure*}[htp]
	\setlength{\abovecaptionskip}{0.2cm}
	\setlength{\belowcaptionskip}{0.0cm}
	\centering
	\includegraphics[height=5.6cm,width=\textwidth]{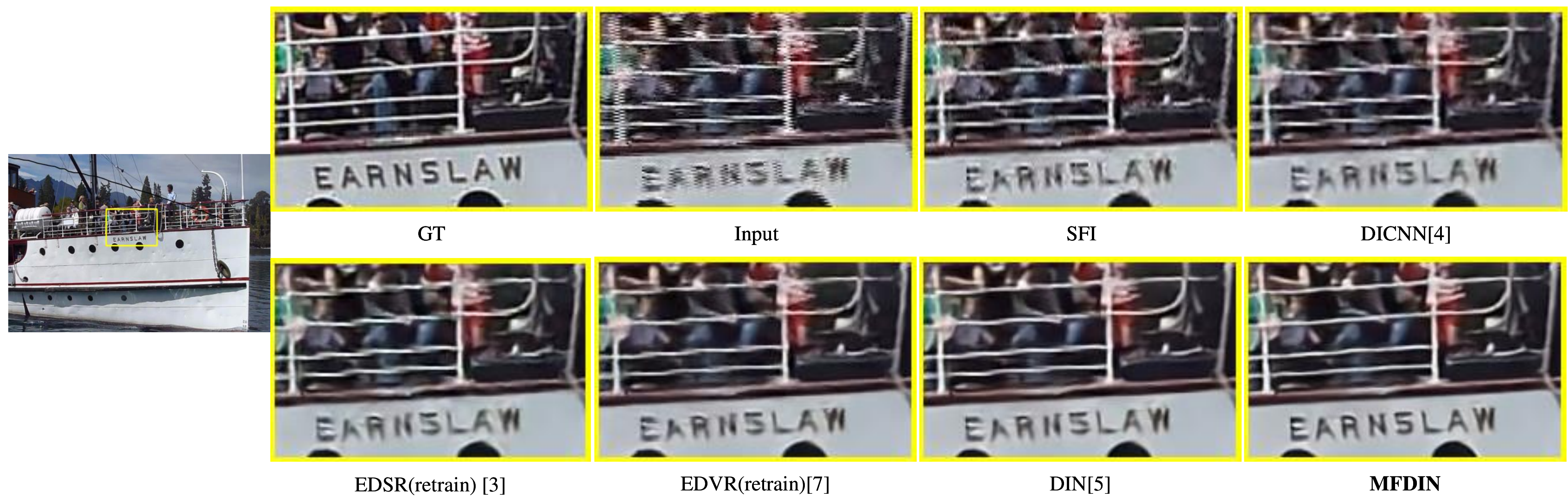}
	\caption{{\color{cblue}Qualitative comparison on the \textbf{Vimeo 90K} synthetic testing set.}}
	\label{fig:11}
\end{figure*}

\begin{table*}[htp]
	\renewcommand{\arraystretch}{1.5}
	\caption{{\color{cblue}Psnr/ssim Results of Different Methods on the Synthetic Datasets}}
	\label{table:3}
	\centering
	\begin{tabular}{c|cccccccccccccc}
		\hline
		& \multicolumn{2}{c}{Input}            & \multicolumn{2}{c}{SFI}              & \multicolumn{2}{c}{DICNN[4]}  & \multicolumn{2}{c}{EDSR[3](retrain)} & \multicolumn{2}{c}{EDVR[6](retrain)} & \multicolumn{2}{c}{DIN[5]}                                                  & \multicolumn{2}{c}{MFDIN}                                                       \\
		\multirow{-2}{*}{} & PSNR              & SSIM             & PSNR              & SSIM             & PSNR         & SSIM        & PSNR            & SSIM            & PSNR            & SSIM            & PSNR                               & SSIM                                & PSNR                                  & SSIM                                   \\ \hline
		YOUKU 2K           & 24.25             & 0.7382           & 27.86             & 0.8439           & 29.16        & 0.8764      & 29.50           & 0.8750          & 31.90           & 0.9036          & {\color[HTML]{00B0F0} \underline{ 32.01}} & {\color[HTML]{00B0F0} \underline{ 0.9071}} & {\color[HTML]{FF0000} \textbf{32.76}} & {\color[HTML]{FF0000} \textbf{0.9132}} \\ \hline
		SJTU 4K            & 21.75             & 0.6971           & 28.16             & 0.8341           & 29.30        & 0.8679      & 30.56           & 0.8887          & 32.42           & 0.8963          & {\color[HTML]{00B0F0} \underline{ 32.52}} & {\color[HTML]{00B0F0} \underline{ 0.8995}} & {\color[HTML]{FF0000} \textbf{33.10}} & {\color[HTML]{FF0000} \textbf{0.9051}} \\ \hline
		Vimeo 90K          & 24.45             & 0.7343           & 28.80             & 0.8381           & 29.87        & 0.8581      & 30.74           & 0.8629          & 31.80           & 0.8781          & {\color[HTML]{00B0F0} \underline{ 32.27}} & {\color[HTML]{00B0F0} \underline{ 0.8813}} & {\color[HTML]{FF0000} \textbf{32.82}} & {\color[HTML]{FF0000} \textbf{0.8880}} \\ \hline
		Parameters         & \multicolumn{2}{c}{\textbackslash{}} & \multicolumn{2}{c}{\textbackslash{}} & \multicolumn{2}{c}{0.05M}  & \multicolumn{2}{c}{1.42M}         & \multicolumn{2}{c}{2.30M}         & \multicolumn{2}{c}{2.22M}                                                & \multicolumn{2}{c}{1.88M}                                                      \\ \hline
		Inference time     & \multicolumn{2}{c}{\textbackslash{}} & \multicolumn{2}{c}{\textbackslash{}} & \multicolumn{2}{c}{0.0164s} & \multicolumn{2}{c}{0.0391s}         & \multicolumn{2}{c}{0.0986s}        & \multicolumn{2}{c}{0.0451s}                                              & \multicolumn{2}{c}{0.0835s}                                                     \\ \hline
		Flops              & \multicolumn{2}{c}{\textbackslash{}} & \multicolumn{2}{c}{\textbackslash{}} & \multicolumn{2}{c}{8.05G}  & \multicolumn{2}{c}{397G}          & \multicolumn{2}{c}{1288G}         & \multicolumn{2}{c}{665G}                                                 & \multicolumn{2}{c}{1375G}                                                      \\ \hline
	\end{tabular}
\end{table*}

\begin{table*}[htp]
	\renewcommand{\arraystretch}{1.5}
	\caption{Quantitative Comparison of Extended Experiments}
	\label{table:4}
	\centering
	\setlength{\tabcolsep}{5.5mm}{
	\begin{tabular}{c|cccccccc}
		\hline
		& \multicolumn{2}{c}{Youku2K}                                  & \multicolumn{2}{c}{SJTU}                                     & \multicolumn{2}{c}{Youku2K}                                  & \multicolumn{2}{c}{SJTU}                                     \\
		& \multicolumn{2}{c}{540i-to-1080P}                            & \multicolumn{2}{c}{1080i-to-4K}                            & \multicolumn{2}{c}{540i(25P)-to-1080p(50P)}                  & \multicolumn{2}{c}{1080i(25P)-to-4K(50P)}                  \\
		& PSNR                         & SSIM                          & PSNR                         & SSIM                          & PSNR                         & SSIM                          & PSNR                         & SSIM                          \\ \hline
		VI+EDSR[3]    & 36.58                        & 0.9520                        & 34.51                        & 0.9366                        & 36.58                        & 0.9522                        & 34.51                        & 0.9370                        \\ \hline
		DICNN[4]+EDSR[3] & 40.29                        & 0.9670                        & 39.05                        & 0.9585                        & \color[HTML]{00B0F0} \underline{40.30}    & \color[HTML]{00B0F0} \underline{0.9670}           & \color[HTML]{00B0F0} \underline{39.04}     & \color[HTML]{00B0F0} \underline{0.9585} \\ \hline
		DIN[5]+SloMo[41]  & \color[HTML]{00B0F0} \underline{40.94} & \color[HTML]{00B0F0} \underline{0.9693}			 & \color[HTML]{00B0F0} \underline{39.71}     & \color[HTML]{00B0F0} \underline{0.9617}    & 35.73                        & 0.9309                        & 33.95                        & 0.9282                        \\ \hline
		MFDIN      & \textbf{\color[HTML]{FE0000} 43.18} & \textbf{\color[HTML]{FE0000} 0.9808} & \textbf{\color[HTML]{FE0000} 42.28} & \textbf{\color[HTML]{FE0000} 0.9747} & \textbf{\color[HTML]{FE0000} 42.04} & \textbf{\color[HTML]{FE0000} 0.9765} & \textbf{\color[HTML]{FE0000} 41.09} & \textbf{\color[HTML]{FE0000} 0.9722} \\ \hline
	\end{tabular}}
\end{table*}

\begin{figure}[htb]
	\centering
	\includegraphics[height=4cm,width=0.32\textwidth]{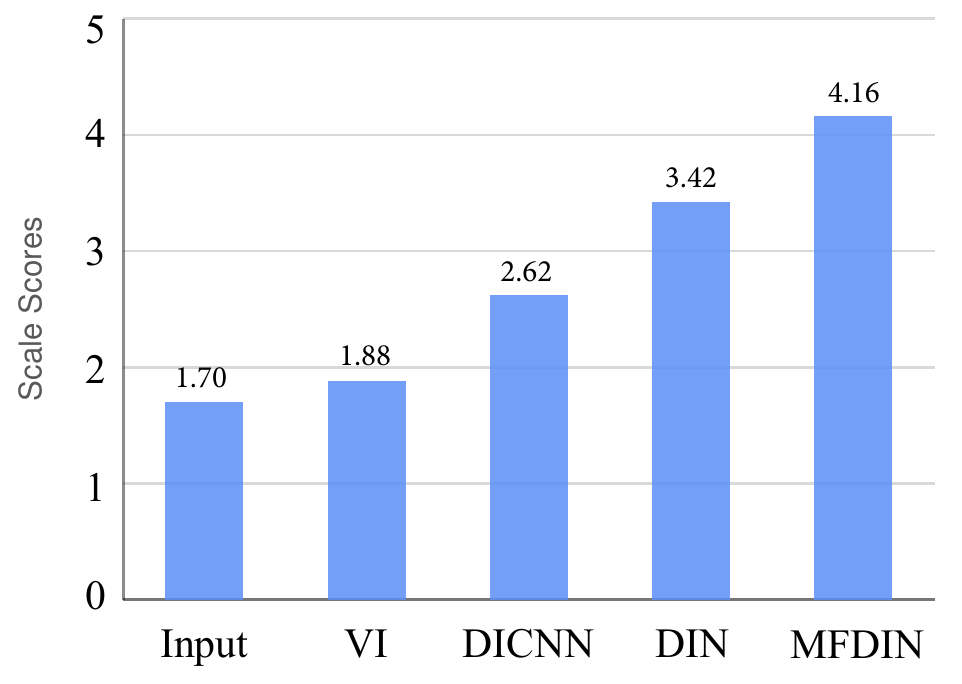}
	\caption{Average SSIS scores of subjective testing on real-world early videos.}
	\label{fig:12}
\end{figure}
\subsection{Results on the synthetic testing sets}
The proposed MFDIN is compared with traditional single field vertical interpolation (SFI) and recent DNN-based deinterlacing methods, \textit{i.e.}, DICNN \cite{zhu2017real} and DIN\cite{zhao2020din}. In addition, we also retrained a deeper single-frame model EDSR \cite{lim2017enhanced} with 16RBs and a SOTA multi-frame model EDVR \cite{wang2019edvr}. Note that their pixel-shuffle (PS) magnification module is removed for deinterlacing task. As in DICNN, these methods take the whole frame as input directly. For fair comparison, all networks are retrained on the same training set. {\color{cblue}Reconstructed results of interlaced and compressed frames on three synthetic datasets are shown in Fig. 9, Fig. 10. and Fig. 11. respectively.} It can be seen that simple SFI method can reduce the comb-teeth effect, but it also enlarges various mixed artifacts. DICNN \cite{zhu2017real} is a lightweight shallow network with few parameters. However, it is only designed for real-time deinterlacing tasks, and thus cannot handle the complex noise in early videos. Because the existing mixed artifacts break the simple inter-field relationship of the interlacing scanning mechanism, the deinterlacing performance of the retrained EDSR is greatly reduced, and various complex noises also cannot be removed well. DIN\cite{zhao2020din} is a deinterlacing method specially designed for early video, which can effectively remove the interlacing comb-teeth and suppress the noise. However, only the information of a single input frame is used, many details cannot be restored totally. In addition, when there is a large motion between frames, as shown in Fig. 10, the unnatural ghost artifacts cannot be eliminated well. Due to the introduction of learning of temporal information, the retrained EDVR can restore more details than the single-frame method. However, under the influence of complex interlacing effects and severe compression noise, inter-frame motion can not be well learned, and ghost artifacts still exist in the restored image. Compared with these SOTA methods, the proposed MFDIN performs much better. On the one hand, fields split and vertical interpolation effectively eliminate the interlaced comb artifacts. On the other hand, alignment and fusion of multi-frames can suppress the ghosting artifacts caused by large-scale motion, and reduce mixed compression artifacts significantly.

{\color{cblue}In addition, Table III lists the quantitative assessments of these methods, which shows that the proposed MFDIN can achieve higher PSNR and SSIM values on three test sets. The number of parameters, the amount of computation and the inference time are also list in Table III, note that the computation and inference time are obtained at a resolution of 720$\times$540.}

\begin{figure*}[htp]
	\centering
	\includegraphics[height=22cm,width=\textwidth]{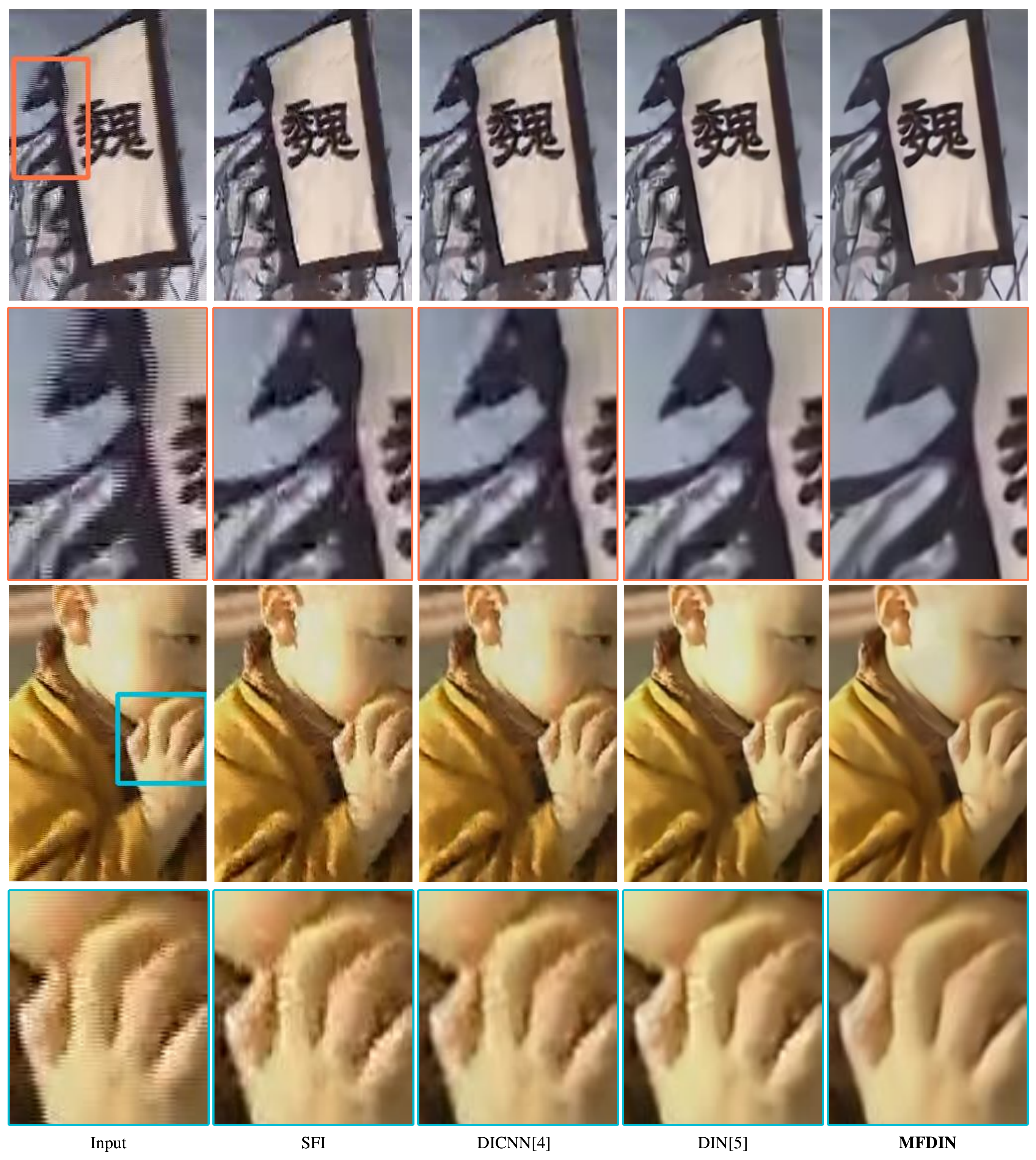}
	\caption{Results on real-world early videos.}
	\label{fig:13}
\end{figure*}
\begin{figure*}[htp]
	\centering
	\includegraphics[height=10.2cm,width=\textwidth]{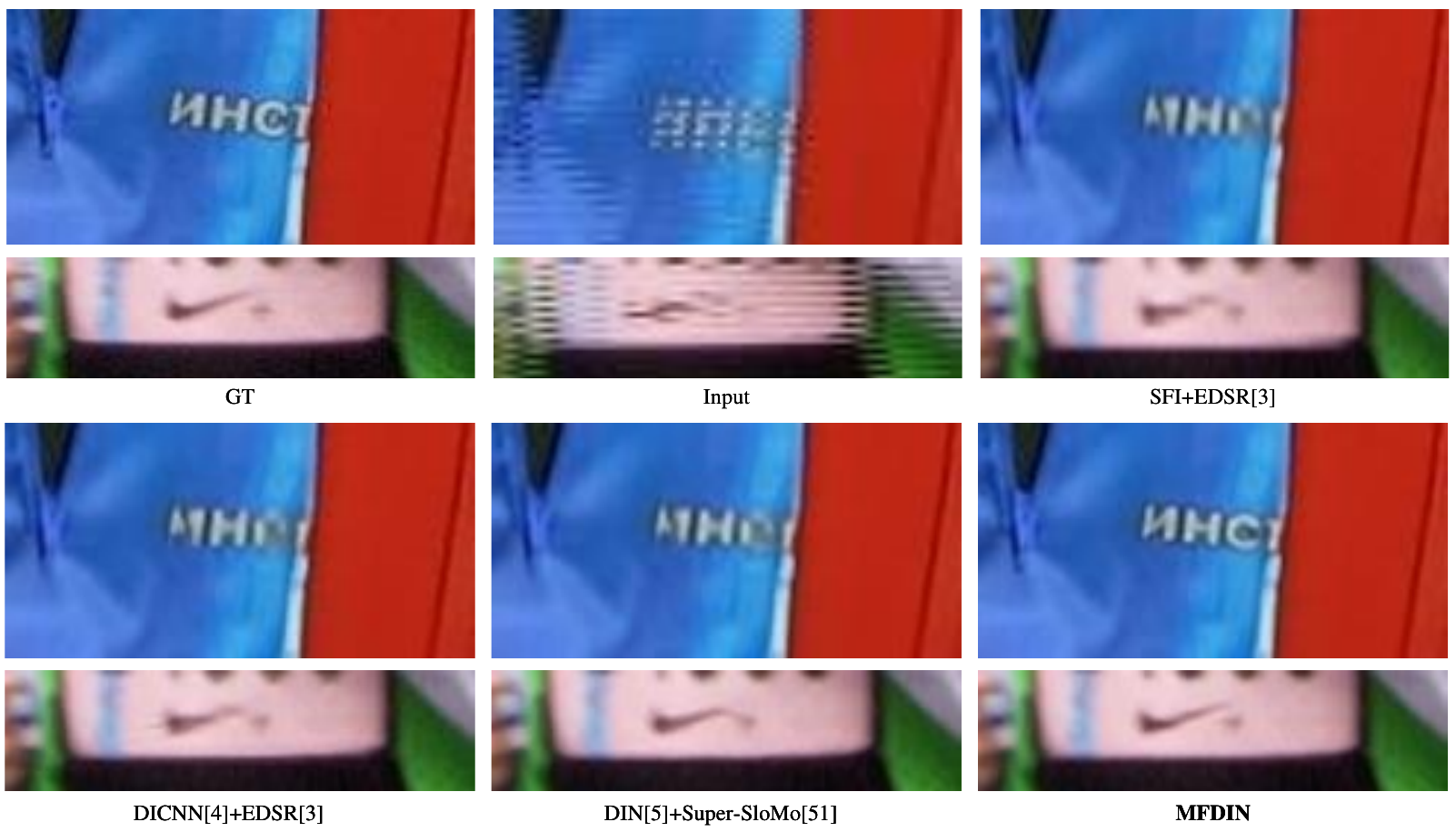}
	\caption{Results of extended experiments on 540i-to-1080p and 1080p-to-4K tasks.}
	\label{fig:14}
\end{figure*}

\subsection{Results on real-world early videos}
Fig. 13 shows the reconstruction results of some early videos. It can be found that there are severe and complex noises such as interlaced comb-teeth, aliasing, noise, and compression artifacts in these early videos. Single-field SFI may amplify various noises in vertical direction. DICNN can reproduce better results than traditional SFI, but it still cannot handle the complex compression noise well. Note that DICNN directly assembles the low-quality input field and reconstructed output field via interlaced scanning. This strategy is not very suitable for early videos because the input frames often contain various noises. DIN can reconstruct high quality results, but it still cannot totally remove severe noise. In contrast, the proposed MFDIN can reconstruct a cleaner and smoother results without obvious artifacts.

Furthermore, subjective single-stimulus-impairment-scale (SSIS) test has been implemented to compare the visual quality of reconstructed early videos. The display condition and test session details were set according to Rec. ITU-R BT.500 \cite{ITU_R_BT500}. A total of 20 observers were invited to score the impairment scales of 5 groups of early video frames, with the rating of 0-5 (5 denotes the highest quality level). The average SSIS scores are illustrated in Fig. 12. It can be found that the MFDIN obtains the highest subjective scores. 

\subsection{Extended experiments}
To verify the scalability and generalization performance of the model, we performed additional experiments on several different application scenarios, such as 540i-to-1080p and 1080i-to-4K. These tasks require deinterlacing, denoising, $2\times$ spatial and temporal interpolation. The proposed MFDIN can be directly applied to these joint tasks by merely adding a PS layer before the last output convolutional layer. In particular, the SFI and DICNN methods do not have the $2\times$ super-resolution ability, while the DIN method does not have the frame rate up-conversion ability. For comparison, we used super-resolution network EDSR \cite{lim2017enhanced} to magnify the results of SFI and DICNN, and utilized video interpolation method Super-SloMo \cite{jiang2018super} to double the frame rate of DIN results.

Quantitative results are shown in Table IV, from which we can find that the proposed multi-frame and multi-stage structure can uniformly deal with the joint and complex tasks, and thus achieves much better performance on these tasks than simple combination of deinterlacing, super-resolution, and frame interpolation methods. In the 540i-to-1080p task, DIN performed better than the two-stage DICNN+EDSR, but the combination of DIN+SloMo did not perform well in the 540i(25P)-to-1080p(50P) task. By contrast, our one-stage method achieves the best performance on both tasks.

The results of different methods on 540i-to-1080p and 1080i-to-4K tasks are shown in Fig. 14. These results illustrate that the MFDIN significantly outperforms other simply combined methods, which can totally remove the interlaced and compression artifacts, restore clearer and more accurate details by means of inter-frame information.

There are some reasons that the designed architecture is suitable for the joint tasks for deinterlacing, compression artifacts removal, super-resolution and frame interpolation. First, the field split and reconstruction strategy can effectively separate the deinterlacing task by using the prior knowledge of interlaced scanning. Second, the backbone of each module is motivated by SOTA super-resolution methods \cite{lim2017enhanced,wang2019edvr,liu2020residual}, which have been proved to be effective for super-resolution and deblocking problems. Third, one interlaced frame contains two fields indeed, which is naturally suitable to double the frame rate. As a result, the MFDIN can even handle these difficult joint problems with a well-designed, efficient and relatively lighter architecture.

{\color{cblue}At last, the shortcomings of the proposed method are discussed. To recover the real-world early videos, MFDIN is designed to suppress the complex interlacing artifacts mixed with other artifacts. However, when there is no other degradations in the videos, the MFDIN is not suitable because some lighter networks can handle normal deinterlacing task, such as DICNN.}


\section{Conclusion}
This paper proposed a multi-frame network (MFDIN) for joint enhancement of early interlacing videos, which can effectively remove complex degradations such as interlacing and compression artifacts. In MFDIN, multiple interlaced frames are firstly split into separate fields, and then feature extraction and vertical interpolation are carried out by a feature extraction module. A temporal alignment module based on deformable convolution was then presented, which can effectively align and fuse the information from multiple fields. Finally, an efficient feature aggregation and reconstruction module was used to further refine details. Experimental results show that the proposed method can reproduce high quality results from early interlaced videos containing severe and complicated unnatural artifacts.


%




%

\bibliographystyle{IEEEtran}
\bibliography{reference}

%

\end{document}